\documentclass[12pt]{iopart}
\usepackage{epsfig}
\DeclareRobustCommand
\openone{\leavevmode\hbox{\small1\normalsize\kern-.33em1}} % identity matrix
\def\pra{Phys.\ Rev.\ A }

\def\prl{Phys.\ Rev.\ Lett.\ }

\begin{document}
\title[D-outcome measurement for a non-locality test]
{D-outcome measurement for a non-locality test}
\author{ W. Son\dag,  Jinhyoung Lee\ddag  and
 M. S. Kim\dag}
\address{\dag\ School of Mathematics and Physics, Queen's University,
  Belfast BT7 1NN, United Kingdom}
\address{\ddag\ Department of Physics, Hanyang University, 17
  Haengdang-Dong, Sungdong-Ku, Seoul, 133-791 Korea}
\ead{w.m.son@am.qub.ac.uk}
\begin{abstract}
For the purpose of the nonlocality test, we propose a general
  correlation observable of two parties by utilizing local $d$-outcome
  measurements with SU($d$) transformations and classical
  communications. Generic symmetries of the SU($d$)
  transformations and correlation observables are found for the test of
  nonlocality. It is shown that these symmetries dramatically reduce
  the number of numerical variables, which is important for
  numerical analysis of nonlocality. A linear combination of the correlation
  observables, which is reduced to the Clauser-Horne-Shimony-Holt (CHSH) Bell's
  inequality for two
  outcome measurements, is led to the Collins-Gisin-Linden-Massar-Popescu (CGLMP)
  nonlocality test for $d$-outcome measurement. As a system to be tested for
  its nonlocality, we
  investigate a continuous-variable (CV) entangled state with $d$
  measurement outcomes. It allows the comparison of nonlocality
  based on different numbers of measurement outcomes on one physical
  system.  In our example of the CV state, we find that a pure entangled
  state of any degree violates Bell's inequality for $d(\ge 2)$
  measurement outcomes when the observables are of SU($d$)
  transformations.
\end{abstract}
\submitto{\JPA}
\maketitle
\section{introduction}

Nonlocality is one of the most profound aspects of a quantum
mechanical system and it is a fundamental resource for quantum
information processing. Nonlocality has been studied commonly in
the operational perspective based on Bell's inequalities for
bipartite two-dimensional systems with dichotomic measurements.
The extensions to arbitrary dimensional systems have been proposed
\cite{gisin,kasz,col}. Recently, Kaszlikowski {\em et al.}
\cite{kasz} considered joint probabilities of two distant
measurements and suggested how to compare the strength of
nonlocality between different dimensional systems for different
numbers of measurement outcomes. A maximally entangled pure system
violates Bell's inequality but as enough white noise is added the
system loses its nonlocality. Kaszlikowski {\em et al.} proposed
the fraction of white noise as a measure of nonlocality, which may
be used for its cross-dimensional comparison. Their numerical
analysis for a maximally entangled state showed that the degree of
violation increases monotonically with respect to the number of
outcomes.

More recently, Collins {\em et al.} \cite{col} developed a family
of Bell's inequalities for an arbitrary finite number of
measurement outcomes. The family of inequalities are in good
agreement with Kaszlikowski {\em et al.}'s results in terms of
their measure of nonlocality. These studies would imply that the
critical fraction of white noise is a useful measure in comparing
the amounts of nonlocality for the different dimensional systems.
The measure of nonlocality based on the noise fraction is, on the
other hand, criticized by Acin {\em et al.} \cite{acin} as they
found that partially entangled systems can give the larger
violation (or stronger resistance to noise) of the Bell's
inequality than the maximally entangled state. The other approach
for a substantial violation of local realism was introduced by van
Dam {\em et. al.} \cite{dam} and they found that CHSH inequality
is the strongest nonlocality test for a bipartite system in terms
of the statistical strength.

In the test of Bell's inequality, a set of unitary transformations
play an important role in the violation of Bell's inequality
because local measurement settings for each party are
characterized by local unitary transformations. In earlier studies
\cite{kasz,col,acin}, the transformations are restricted to
quantum Fourier transformations (QFT). For a $d$-dimensional
system the most general unitary transformation forms the group
SU($d$). It is thus questionable whether the QFT is sufficient to
fully reveal quantum nonlocality. It has been known that the QFT
is sufficient for a maximally entangled system of $d=2~
\mbox{and}~ 3$ \cite{kasz}. However, it is still an open question
for other dimensional systems and, more importantly, for an
arbitrarily entangled system. This question is investigated in
this paper.

Quantum nonlocality for continuous-variable (CV) systems has been
studied in various contexts. Bell argued that the original
Einstein-Podolsky-Rosen (EPR) state \cite{EPR} would not violate
Bell-like inequalities since it has a positive-definite Wigner
function and thus its correlation function with respect to
position and momentum observables can be simulated by local hidden
variables. On the other hand, introducing dichotomic measurements
such as even or odd parities of the photon number and presence or
absence of photons, Banaszek and W\'{o}dkiewicz
\cite{banaszek,jeong00} showed the nonlocality of the EPR state.
The measurements follow displacement operations, that is,
translations in the phase spaces of the modes. However, the scheme
by Banaszek and W\'{o}dkiewicz did not give a maximal violation
for the inequality. This motivated Chen {\em et al.} \cite{chen}
to investigate another type of observable, with the unitary
transformations other than the displacement in phase space, which
results in the maximal violation for EPR state. The observable,
so-called ``pseudo-spin'' operator, is defined as tensor
summations of Pauli spin operators, which is element of an SU($2$)
group.

The generalization of a dichotomic measurement to an arbitrary
finite number of outcomes for a nonlocality test of a CV state was
proposed by Brukner {\em et al.} \cite{caslav} as establishing a
correspondence between a CV and a discrete system of an arbitrary
finite dimension. However, in their work, it is not clear whether
the correspondence can be given as a physically plausible map,
{\it i.e.} a completely positive (CP) map. Moreover, in their
analysis for nonlocality, Brukner {\em et al.} did not employ
general transformations in SU($d$) but the simple QFT
transformations in varying the configuration of measurements. Thus
the question of QFT being sufficient to reveal nonlocality arises
in CV systems as well as in the finite-dimensional systems.

In this paper, we formulate the Bell's inequality in terms of a
linear summation of correlation functions which utilize the most
general projective $d$-outcome measurements. For the correlation
function, we introduce a general form of correlation observable
between two $d$-level systems and find the eigenvalues from the
generic conditions that any correlation function should satisfy.
For the observable, we exploit all the possible unitary
transformations in the SU($d$) group on the configurations of
local $d$-outcome measurements. The subgroup algebra of SU($d$)
allows us to prove that $(d^2-d)$ number of real parameters are
sufficient to describe the local unitary operation. By inspecting
symmetries and performing numerical analysis, we show, while the
QFT suffices for a maximally entangled system, a partially
entangled system requires more general transformations in order to
fully investigate its nonlocality.

The Bell's inequalities are applied to a CV state whose
infinite-dimensional Hilbert space is decomposed onto the tensor sum of
$d$-dimensional subspaces. The decomposition maps any CV state onto an
arbitrary finite-dimensional state. We prove that the mapping is
linear, trace preserving, complete positive. After applying the mapping,
we investigate the violation of Bell's inequality for the two-mode
squeezed vacuum state, as an example of a CV state, with the different
outcome measurements. To search for the optimal violation in SU($d$)
transformation, several numerical methods are assessed.

\section{Bell's inequalities with $d$-outcome measurements}
In this section, we investigate Bell's inequalities by considering
the SU($d$) group of transformations for the measurement with $d$
outcomes. The series of inequalities may be derived by introducing
``classically correlated observables'' which can be constructed by
local measurements and classical communications.

\subsection{Special unitary transformation for the $d$-outcome measurement}
A measurement of a system is represented by a Hermitian operator which
is called an observable. Any Hermitian operator on a $d$-dimensional
Hilbert space ${\cal H}_d$ can be expanded by the identity operator and
the group generators of SU($d$) algebra.  Such a typical description in
terms of group generators was introduced by Hioe and Eberly \cite{hioe}.
In order to obtain the generators of the SU($d$) group, one may
introduce transition-projection operators
\begin{equation}
\label{eq:basis}
\hat{P}_{jk}=|j\rangle\langle k|,
\end{equation}
where $\{|j\rangle\}$ is an orthonormal basis set on ${\cal H}_d$.  Now,
the $(d^2-1)$ Hermitian operators are constructed as
\begin{eqnarray}
\hat{u}_{jk}&=&\hat{P}_{jk}+\hat{P}_{kj}\\
\hat{v}_{jk}&=&i(\hat{P}_{jk}-\hat{P}_{kj})\\
\hat{w}_{l}&=&-\sqrt{\frac{2}{l(l+1)}}\left(\sum_{i=1}^{l}
\hat{P}_{ii}-l\hat{P}_{l+1 l+1}\right)
\end{eqnarray}
where $1\leq l \leq d-1$ and $1 \leq j < k \leq d$. It is easy to check
that when $d=2$ these generators are Pauli spin operators.

The set of
$G=\{\hat{u}_{12},\hat{u}_{13},\ldots,\hat{v}_{12},\hat{v}_{13},
\ldots,\hat{w}_{1},\ldots,\hat{w}_{d-1}\}$ is composed of generators for
SU($d$) group, fulfilling the relations of tracelessness
$\mbox{Tr}(\hat{s}_j)=0$ and orthogonality $\mbox{Tr}(\hat{s}_i
\hat{s}_j)= 2 \delta_{ij}$ for $\hat{s}_i, \hat{s}_j \in G$. The
elements $\hat{s}_i \in G$ hold the algebraic relation,
\begin{equation}
\label{eq:alge}
[\hat{s}_j,\hat{s}_k]= 2 i \sum_l f_{jkl} \hat{s}_l
\end{equation}
where $f_{jkl}$ is the antisymmetric structure constant of the SU($d$)
algebra.

The set $G$ can be divided into the three mutually exclusive subsets of
operators: $U=\{\hat{u}_{jk}\}$, $V=\{\hat{v}_{jk}\}$, and $W
=\{\hat{w}_{l}\}$, which contain $d(d-1)/2$, $d(d-1)/2$, and $(d-1)$
elements respectively. The operators in each subset satisfy the
algebraic relations
\begin{eqnarray}
\label{eq:commu} \left[\hat{u}_{ij},\hat{u}_{kl}\right]&=&-i\{
\delta_{jl}\left(1-\delta_{ik}\right)\hat{v}_{ik}+
\delta_{ik}\left(1-\delta_{jl}\right)\hat{v}_{jl}\nonumber\\
&&~+\delta_{jk}\left(1-\delta_{il}\right)\hat{v}_{il}-
\delta_{il}\left(1-\delta_{kj}\right)\hat{v}_{kj}
\},~~\nonumber\\
\left[\hat{v}_{ij},\hat{v}_{kl}\right]&=&-i\{\delta_{jl}\left(1-\delta_{ik}\right)
\hat{v}_{ik}+\delta_{ik}\left(1-\delta_{jl}\right)\hat{v}_{jl}\\
&&~-\delta_{jk}\left(1-\delta_{il}\right)\hat{v}_{il}+
\delta_{il}\left(1-\delta_{kj}\right)\hat{v}_{kj}
\},~~\nonumber\\
\left[\hat{w}_{i},\hat{w}_{j}\right]&=&0.\nonumber
\end{eqnarray}
The commutators between the elements from $U$ or $V$ result in the
operators in the subset $V$ while any elements in $W$ commute to
each other. The commutation relations among the operators from
the different subsets can be found as
\begin{eqnarray}
\label{eq:commu1}
\left[\hat{u}_{ij},\hat{w}_{l}\right]&=&-(x_{il}-x_{jl})\hat{v}_{ij}\nonumber\\
\left[\hat{v}_{ij},\hat{w}_{l}\right]&=&-i(x_{il}-x_{jl})\hat{u}_{ij}\\
\left[\hat{u}_{ij},\hat{v}_{kl}\right]&=&i(\delta_{jk}\hat{u}_{il}+\delta_{ik}\hat{u}_{jl}-
\delta_{jl}\hat{u}_{ik}-\delta_{il}\hat{u}_{kj})\nonumber\\
&~&+2i\delta_{ik}\delta_{jl}\left(|j\rangle\langle
j|-|i\rangle\langle i|\right)\nonumber
\end{eqnarray}
where the coefficient $x_{il}$ is given by
\begin{equation}
x_{il}=-\sqrt{\frac{2}{l(l+1)}}\left(\sum_{k=1}^{l}\delta_{ik}-l\delta_{i l+1}\right)
\end{equation}
and $\delta$ is the Kronecker delta function. As a summary, one may
symbolically express the commutation relation among the subsets $U$,
$V$, and $W$ as
\begin{eqnarray}
\label{eq:commu2}
\left[U,U\right]\propto V,~~~ \left[V,V\right]&\propto& V,
~~~\left[W,W\right]\propto 0 \nonumber\\
\left[U,W\right]\propto V,
~~~\left[V,W\right]&\propto& U,
~~~\left[U,V\right]\propto U+W.
\end{eqnarray}

Using the SU($d$) group generators, any Hermitian operator on the
$d$-dimensional Hilbert space is represented by
\begin{equation}
\label{eq:obser}
\hat{\Omega}(\vec{a})=\frac{a_0}{d}\hat{\openone}+\frac{1}{2}\sum_{j=1}^{d^2-1}a_j
\hat{s}_j,
\end{equation}
where $a_0=\mbox{Tr}\hat{\Omega}(\vec{a})$ and $a_j=\mbox{Tr} \hat{s}_j
\hat{\Omega}(\vec{a})$ are real numbers due to the hermiticity condition
of the observable $\hat{\Omega}(\vec{a})$.  The coefficient $a_j$
comprises a ($d^2-1$)-dimensional vector
$\vec{a}=(a_1,\ldots,a_{d^2-1})$ which we call a {\em generalized Bloch
  vector}, while $a_0$ is constant over any SU($d$) transformations.

In the Heisenberg picture, the unitary transformation $\hat{U}$ of the
Hermitian operator,
\begin{equation}
\label{eq:transform}
\hat{\Omega}(\vec{a})\rightarrow\hat{\Omega}(\vec{a}')=\hat{U}
\hat{\Omega}(\vec{a})\hat{U}^{\dagger},
\end{equation}
can also be described as a transformation of the generalized Bloch
vector $\vec{a}$. Decomposing $\hat{\Omega}(\vec{a}')$ in the form
(\ref{eq:obser}) with coefficients $a_j '$ and using the invariance of
the trace under cyclic permutation, the components $a_j '$ of the
transformed generalized Bloch vector are found to be
\begin{equation}
a_j'=\mbox{Tr}(\hat{s}_j
\hat{\Omega}(\vec{a}'))=\mbox{Tr}(\hat{U}^{\dagger}\hat{s}_j\hat{U}
\hat{\Omega}(\vec{a})).
\end{equation}
Since $\hat{U}^{\dagger}\hat{s}_j\hat{U}$ is also Hermitian and
traceless, it can be expanded in terms of the SU($d$) generators as
\begin{equation}
\label{eq:trans} \hat{U}^{\dagger}\hat{s}_j\hat{U} \equiv \sum_{k}
T_{jk} \hat{s}_k
\end{equation}
where
$T_{jk}=\frac{1}{2}\mbox{Tr}(\hat{U}^{\dagger}\hat{s}_j\hat{U}\hat{s}_k)$
is an element of a $(d^2-1)\times(d^2-1)$ real matrix. The matrix $T$
represents the direct relation between the transformed and untransformed
generalized Bloch vectors $a_k'=\sum_{j}T_{jk}a_j$.  As the norm of the
generalized Bloch vector remains constant under the transformation, the
real matrix $T$ is orthogonal.

An operator $\hat{U}$ in SU($d$) can be represented in terms of the
group generators $\vec{s}=(\hat{s}_1, \hat{s}_2,\cdots,
\hat{s}_{d^2-1})$ as
\begin{equation}
\label{eq:untary}
\hat{U}(\vec{p})=\exp(-i\vec{p}\cdot\vec{s})
\end{equation}
where $\vec{p}$ is a $(d^2-1)$-dimensional parameter vector. The
parameterization in Eq.~(\ref{eq:untary}) is said to be canonical.
Experimentally, for the optical device, it is possible to realize
the discrete unitary operation in SU($d$) using biased multiport
beam splitters \cite{reck}. In order to derive the explicit
matrix elements of $T$ in Eq.~(\ref{eq:trans}) corresponding to
the unitary operator $\hat{U}(\vec{p})$, one may consider a set
of differential equations for the generators:
\begin{eqnarray}
\label{eq:partial}
\frac{\partial}{\partial t}\hat{s}_j(t)
&=&\hat{U}^{\dagger}(t\vec{p})\left\{i \sum_{k}p_k \left[\hat{s}_k,
\hat{s}_j\right]\right\}\hat{U}(t\vec{p})\nonumber\\
&=&\sum_l \left(-2\sum_{k} p_k f_{kjl}\right)\hat{s}_l(t)
\end{eqnarray}
where
$\hat{s}_j(t)=\hat{U}^{\dagger}(t\vec{p})\hat{s}_j\hat{U}(t\vec{p})$.
After solving the differential equation and setting $t=1$, the matrix
$T$ is derived in terms of the parameter vector $\vec{p}$ and the
antisymmetric structure constant $f_{kjl}$ as
\begin{equation}
\label{eq:tmat}
T(\vec{p})=\exp(-2 F(\vec{p})),~~~~\mbox{where}
~~~F_{jl}(\vec{p})=\sum_{k}p_k f_{kjl}.
\end{equation}
The antisymmetric characteristics of the structure constant $f_{kjl}$ is
related with the orthogonality of $T$ as
$T^{T}T=TT^{T}=\openone_{d^2-1}$ where $\openone_{d^2-1}$ is an identity
matrix on $(d^2-1)$-dimensional vector space.

It is notable that a commutation relation appears in
Eq.~(\ref{eq:partial}). From the fact that the group generators
$\{\hat{s}_j\}$ are divided into three subsets $U$, $V$ and $W$ and the
generators of each subset satisfy the algebraic relations
(\ref{eq:commu}) and (\ref{eq:commu1}), one can find some symmetries for
the rotation of generators. Especially, since $[\hat{w}_i,\hat{w}_j]=0$,
it is possible to find that the rotation of generators $\{\hat{w}_l\}\in
W$ along the direction of any $\hat{w}_j$ results in the generator
itself as
\begin{equation}
\label{eq:rot} \exp(-i p_{w_j}\hat{w}_j)\hat{w}_l \exp(i
p_{w_j}\hat{w}_j)=\hat{w}_l
\end{equation}
where $p_{w_j}$ is the $\hat{w}_j$ component of the parameter vector
appeared in Eq.~(\ref{eq:untary}).

Eq.~(\ref{eq:rot}) implies that the dimensionality of the nontrivial
parameter vector $\vec{p}$ for the unitary transformation of the
Hermitian operator (\ref{eq:transform}) can be reduced to $(d^2-d)$.
Without loss of generality, the Hermitian operator (\ref{eq:obser}) can
be written with a given orthogonal basis $\{|j\rangle\}$ and the unitary
operator of the basis transformation as
\begin{equation}
\label{eq:obser2}
\hat{\Omega}(\vec{a})=\hat{U}(\vec{p}_a)\sum_{j=1}^{d}
\Omega_j|j\rangle\langle j| ~\hat{U}^{\dagger}(\vec{p}_a)
\end{equation}
where $\Omega_j$ is the non-degenerate eigenvalue. The generators
in the subset $W$ are sufficient to reconstruct all the diagonal
bases $\{|j\rangle\langle j|\}$, that is,
\begin{equation}
\label{eq:recon} |j\rangle\langle j|=\frac{1}{d}\hat{\openone}-
\sum_{k=0}^{d-j} g_k^j \hat{w}_{j-1+k}
\end{equation}
where $\hat{g}_k^j=(1-j \delta_{k 0})\sqrt{\frac{1}{2
(j+k)(j+k-1)}}$. With help of Eqs.~(\ref{eq:rot}) and
(\ref{eq:recon}), one can see that the dimensionality of the
nontrivial parameter vector $\vec{p}_a$ in Eq.~(\ref{eq:obser2})
is $(d^2-d)$. This implies that \textit{any unitary
transformation in SU($d$) for the observable in the
$d$-dimensional Hilbert space is sufficient with $(d^2-d)$ number
of real parameters.}

\subsection{Classically correlated observable with $d$-outcome measurement}
Two observers, say, Alice and Bob perform local measurements on
their own $d$-level systems and they communicate their outcomes
via a classical channel. A classically correlated observable is
thus constructed as assigning a weight $\mu_{i,j}$ for the pair of
outcomes, $i$ and $j$:
\begin{equation}
\label{eq:cco} \hat{E}=\sum_{i,j=1}^{d}\mu_{i,j}
|i\rangle_a\langle i|\otimes|j\rangle_b\langle j|.
\end{equation}
The correlation coefficient matrix $\mathbf{\mu}$ is a $d\times d$ real
matrix. It is notable that Eq.(\ref{eq:cco}) is the most general form of
a correlation measure between any two $d$-level systems.  The
correlation observable $\hat E$ involves $d^2$ correlation coefficients.
We show that without loss of generality, the number of independent
parameters reduces to $d$ and we determine their values.

We require that a correlation observable should satisfy the following
conditions.
\begin{itemize}
\item[C.1] A correlation function should be indifferent to local
  polarization, which means that
\begin{equation}
\mbox{Tr}\hat{E}\hat{\rho}_A\otimes\openone_B =
\mbox{Tr}\hat{E}\openone_A\otimes\hat{\rho}_B=0,
\label{eq:condition1}
\end{equation}
where $\hat{\rho}_{A,B}=\mbox{Tr}_{B,A} \hat{\rho}$ are the reduced
density operators.  This raises the following condition:
\begin{equation}
\label{eq:c1} \sum_{j}\mu_{i,j} =0,~~\forall i
~~~\mbox{and}~~~\sum_{i}\mu_{i,j} =0,~~\forall j.
\end{equation}
\end{itemize}
For the case of two outcomes, there is the well accepted
correlation matrix
$\mathbf{\mu}=\{\{\mu_{1,1},\mu_{1,2}\},\{\mu_{2,1},\mu_{2,2}\}\}=
\{\{1,-1\},\{-1,1\}\}$.  Here, the translational symmetry,
$\mu_{1,1}=\mu_{2,2}$ and $\mu_{12}=\mu_{21}$, and equal spacing,
$\mu_{1,1}-\mu_{2,1}=2$ leads the correlation observable to
optimize the measure of correlation.  We generalize these in the
following two conditions.
\begin{itemize}
\item[C.2] The correlation coefficients are unbiased over their outcomes
  (translational symmetry within modulo $d$):
\begin{equation}
\label{eq:c2} \mu_{i+k, j+k}=\mu_{i,j},~~~\forall k.
\end{equation}
\item[C.3] The coefficients are equally separated and normalized
  (maximal discrimination):
\begin{equation}
\mu_{i,j}-\mu_{i+1, j}=\frac{2}{d-1},~~~ \mbox{for}~~ i\geq j
\end{equation}
\end{itemize}
The condition C.2 leads the correlation matrix $\mu$ to be in the form
of
\begin{equation}
\label{eq:mu}
\mu =
\left(
\begin{array}{ccccc}
  \mu_1&\mu_2&\mu_3&\cdots&\mu_{d}\\
  \mu_{d}&\mu_1&\mu_2&\cdots&\mu_{d-1}\\
  \vdots&\vdots&\vdots&\ddots&\vdots\\
  \mu_2&\mu_3&\mu_4&\cdots&\mu_{1}
\end{array}
\right)
\end{equation}
and further the condition C.1 implies that
\begin{equation}
\sum_l \mu_l=0.
\end{equation}
The condition C.3 determines all the $\mu_l$'s such that $\mu_1=1$ for a
maximally correlated state, $\mu_d=-1$ for a maximally anti-correlated
state and the other $\mu_l$'s are assigned to have equally spaced values
between 1 and -1. Thus, the three conditions C.1, C.2, and C.3 uniquely
determine the correlation matrix $\mu$,
\begin{equation}
\label{eq:eigen} \mu_{i,j}=1-2\frac{(i-j) \mbox{mod} ~d}{d-1}.
\end{equation}

Using Eq.~(\ref{eq:recon}), the correlation observable $\hat{E}$ in
Eq.~(\ref{eq:cco}) can be written in terms of the SU($d$) generators as
\begin{equation}
\label{eq:corr3}
\hat{E}=\sum_{k,l=1}^{d-1}\tilde{\mu}_{k,l}\hat{w}_k\otimes\hat{w}_{l}
\end{equation}
where $\tilde{\mu}_{k,l}$ is the transformed correlation matrix
from $\mu_{j,k}$. That is,
\begin{equation}
\tilde{\mu}_{k,l}=\sum_{i=1}^{k+1}\sum_{j=1}^{l+1}g^i_{k-i+1
}g^j_{l-j+1} \mu_{i,j}
\end{equation}
where $g^i_{k}$ is given in Eq.~(\ref{eq:recon}). Note that the
correlation observable $\hat{E}$ in Eq.~(\ref{eq:corr3}) does not
contain any local identity operator $\openone_d$ due to the condition
C.1.  Further, the observable transformed by local unitary operations is
written as
\begin{eqnarray}
\label{eq:corr4}
\hat{E}(\vec{p},\vec{q})&=&\hat{U}(\vec{p})\otimes\hat{U}(\vec{q})
\hat{E}\hat{U}^{\dag}(\vec{p})\otimes\hat{U}^{\dag}(\vec{q})\\
&=&\sum_{l,m=1}^{d^2-1} \tilde{\mu}_{l,m}(\vec{p},\vec{q})
\hat{s}_{l}\otimes\hat{s}_{m}\nonumber
\end{eqnarray}
where
$\tilde{\mu}(\vec{p},\vec{q})=T^{T}(\vec{p})\tilde{\mu}T(\vec{q})$.  The
unitary operators $\hat{U}(\vec{p})$ and $\hat{U}(\vec{q})$ determine
the measurement configuration for each side. Without any constraint for
the $d$-outcome measurement, the unitary operators are subjected to the
SU($d$) group.
%which was defined with the
%$(d^2-d)$ free parameters as seen in Eq.~(\ref{eq:obser2}).

\subsection{Bell's inequalities for bipartite $d$-dimensional system}

In order to investigate nonlocality of a bipartite system, we introduce
a Bell function which can be constructed by a linear combination of
correlation functions of two parties. The Bell function can be written
without loss of generality as
\begin{equation}
\label{eq:bell1} B=\sum_{i} c_{i} E(\vec{p}_i,\vec{q}_i)
\end{equation}
where the correlation function
$E(\vec{p},\vec{q})=\mbox{Tr}\hat{E}(\vec{p},\vec{q})\hat{\rho}$ and
$\vec{c}=\{c_{i}\}$ is an arbitrary vector which satisfies a normalized
condition $\sum_i c_i = 2$ to make the Bell function $B$ a polytope
\cite{werner}. Note that the correlation function $E(\vec{p},\vec{q})
\in [-1,1]$ for all $\vec{p}$ and $\vec{q}$.  The classically correlated
observable $\hat{E}(\vec{p},\vec{q})$ can be written as
\begin{equation}
\label{eq:corr1} \hat{E}(\vec{p},\vec{q})= \sum_{i,j}\mu_{i,j}
\hat{P}_i(\vec{p})\otimes\hat{P}_j(\vec{q}),
\end{equation}
where the projector
$\hat{P}_i(\vec{p})=\hat{U}(\vec{p})|i\rangle\langle
i|\hat{U}^{\dagger}(\vec{p})$ is for the $i$-th outcome with the
measurement configuration $\vec{p}$ and the correlation matrix
$\mu$ is given in Eq.~(\ref{eq:eigen}).  The joint probability
that Alice and Bob obtain the outcomes $i$ and $j$ with the
measurement configurations $\vec{p}$ and $\vec{q}$ is given by
\begin{equation}
P_{ij}(\vec{p},\vec{q})=\mbox{Tr}\left(\hat{P}_i(\vec{p})
\otimes\hat{P}_j(\vec{q})\hat{\rho}\right).
\end{equation}
This implies that, from the joint probabilities for a given measurement,
one can obtain the correlation functions for different measurement
configurations and thus the Bell function (\ref{eq:bell1}).

A Bell function has its boundary which is allowed by a local
realistic model. It is worthwhile mentioning that
quantum-mechanically correlated states do not violate the
boundaries of all the possible Bell functions (\ref{eq:bell1}).
Only the Bell functions whose boundaries are violated by
quantum-mechanically correlated states are of interest in the
test of nonlocality \cite{massar}. In this paper, we do not try
to find all the classical boundaries. Instead, we consider the
Bell function whose classical upper bound is 2 with the
particular vector $\vec{c}=(1,1,1,-1)$,
\begin{equation}
\label{eq:bell}
B= E(\vec{A}_1,\vec{B}_1) + E(\vec{A}_2,\vec{B}_2)
+ E(\vec{B}_2,\vec{A}_1) - E(\vec{A}_2,\vec{B}_1).
\end{equation}
After a little algebra, one realizes that the Bell function
(\ref{eq:bell}) is exactly the same as the Bell function of
Collins-Gisin-Linden-Massar-Popescu (CGLMP) \cite{col}, whose
classical bounds are found as 2 with help of joint probabilities.

Note that the third term of the correlation function in
Eq.~(\ref{eq:bell}) has the parameters for the measurement
configurations exchanged. In general, the correlation function
$E(\vec{p}, \vec{q})$ depends on exchanging the parameter vectors,
\begin{equation}
E(\vec{p}, \vec{q})\neq E(\vec{q}, \vec{p}).
\end{equation}
The correlation function is invariant for the parameter exchange only
for dichotomic measurements in which case the Bell function
(\ref{eq:bell}) is led to the CHSH Bell's inequality \cite{clauser}.

In order to find the quantum mechanical maximum for the Bell
function Eq.~(\ref{eq:bell}), CGLMP used the QFT unitary
transformation,
\begin{equation}
\hat{U}_{QFT}(\vec{A})= \frac{1}{\sqrt{d}}\sum_{j,k}
e^{i\frac{2\pi}{d}j(k+\phi_A)}|j\rangle\langle k|,
\end{equation}
which has only a single parameter $\phi_A$ to be adjusted for the
measurement configuration. They found that for the
$d$-dimensional {\em maximally} entangled state their Bell
function, which is the same as Eq.~(\ref{eq:bell}), has its
maximum
\begin{equation} \label{eq:collins}
B_d = 4d \sum_{l=0}^{d-1}\left(1-\frac{2l}{d-1}\right) \frac{1}{2 d^3
\sin^2[\pi(l+1/4)/d]}
\end{equation}
when $(\phi_{A_1},\phi_{A_2},\phi_{B_1},\phi_{B_2})=(0,1/2,1/4,-1/4)$.
This is always larger than the local realistic upper bound 2 and
increases as the number, $d$, of measurement outcomes increases.
However, as a special subset of the unitary group U($d$), it is unclear
whether the QFT measurement is optimal for the test of nonlocality when
$d>3$ even though it has been known that this is the case for maximally
entangled states of $d=2$ and $d=3$ \cite{kasz}.

\begin{figure}[htbp]
\begin{center}
\leavevmode
   \epsfxsize=11 cm
   \epsfysize=9 cm
  \epsfbox{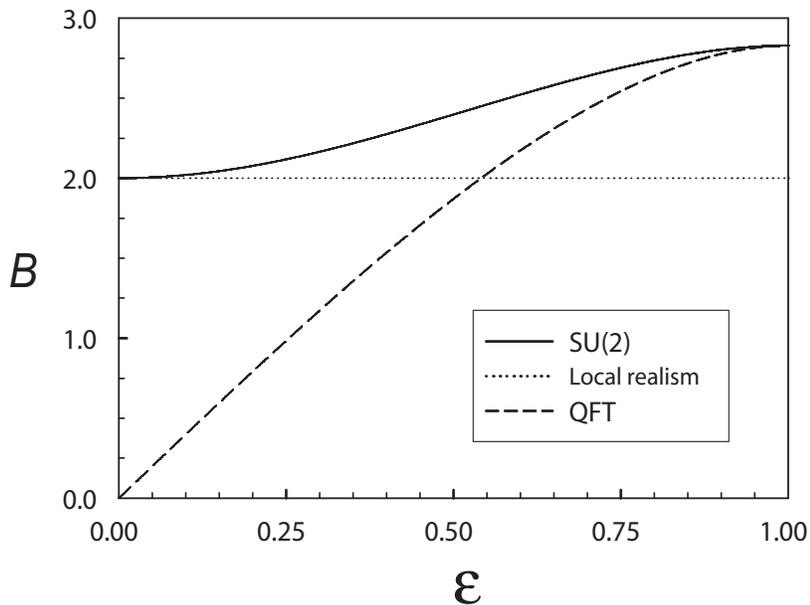}
\end{center}
 \caption{\label{fig1}Bell function $B$ with respect to the entanglement
    $\varepsilon=\sqrt{2} \sin \varphi$ for a two-dimensional bipartite state.
    Solid line represents for the case of SU(2) measurement and dashed
    line for the measurement with QFT.}
\end{figure}

The raised question becomes rather dramatic if a state is
partially entangled.  For example, when $d=2$, the violation of
the Bell's inequality in Eq.~(\ref{eq:bell1}) is plotted in
Fig.~\ref{fig1}. Note that the Bell's inequality becomes the CHSH
Bell's inequality when $d=2$. The state is assumed to be in a pure
state of
$|\psi\rangle=\cos\varphi|00\rangle+\sin\varphi|11\rangle$. The
Bell functions are optimized for the different measurement
configurations : the dashed line is obtained by the QFT and the
solid line by the SU($2$) transformations. The figure shows that
the QFT is not an optimal transformation in revealing the
nonlocality of the partially entangled state.  It is required to
consider the general SU($d$) transformations for the optimal
nonlocality test when a state is in a partially entangled state.

\vspace{0.5cm}

\section{Highly degenerate measurement for a CV state}

In this section, we consider a CV state as a system to test its
nonlocality. For the purpose of the nonlocality test, one needs to
introduce a proper measurement which can show the violation of any
Bell's inequality. Generally, for the case of a CV state, the
spectrum of its measurement is continuous and the number of
non-degenerate eigenvalue is infinite. Therefore, a difficulty
arises, even in principle, in measuring the infinite number of
outcomes and the test of its nonlocality. Several possible methods
which can overcome such a difficulty have been suggested
\cite{banaszek,chen,caslav}. These methods have adopted
measurements which give a finite number of outcomes from a CV
state. This measurement naturally assumes an infinite degeneracy
in the measurement. Recently, it is also found that homodyne
measurement after single photon substraction from a CV state can
play an essential role for a loophole-free nonlocality test
\cite{garcia-patron}.

In this section, we formulate the explicit form of an observable
$\hat{A}$ which can give a finite $d$-number of outcomes from the
measurement on a CV system. The observable corresponds to a
mapping from a CV state to an arbitrary $d$-dimensional system and
the mapping is the same mapping which was suggested by Brukner
\textit{et al.}\cite{caslav}. We show, here, that the mapping is a
linear, trace preserving, and complete positive (CP) map which
implies that the density matrix of a CV state can be legitimately
transformed into a finite dimensional state.

\subsection{$d$-outcome measurement for a CV system}

An observable $\hat{A}(\vec{a})$ which gives $d$ outcomes from the
measurement on a CV state can be found as a direct sum of the infinite
number of the $d$-dimensional observables;
\begin{equation}
\hat{A}(\vec{a})= \left(
\begin{array}{cccc}
\hat{\Omega}(\vec{a})&0&0&\cdots\\
0&\hat{\Omega}(\vec{a})&0&\cdots\\
0&0&\hat{\Omega}(\vec{a})&\cdots\\
\vdots&\vdots&\vdots&\ddots\\
\end{array}
\right)
\end{equation}
where $\hat{\Omega}(\vec{a})$ is the observable which is in the
$d$-dimensional Hilbert space and has the explicit form as was given in
Eq.~(\ref{eq:obser}). The measurement with the observable
$\hat{A}(\vec{a})$ produces infinite degeneracy in each outcome since it
counts every $d$ modulo basis state as the same outcome.  Alternatively,
the observable $\hat{A}(\vec{a})$ can be written as
\begin{equation}
\hat{A}(\vec{a})=\sum_{m=0}^{\infty}\sum_{jk=0}^{d-1}
\Omega_{jk}(\vec{a}) |dm+j\rangle\langle dm+k|
\end{equation}
where $\Omega_{jk}(\vec{a})$ is the matrix element of the
$d$-dimensional observable which is parameterized with the $d^2-1$
dimensional generalized Bloch vector $\vec{a}$.

With the observable $\hat{A}(\vec{a})$, one can establish the
mapping between a CV state and an arbitrary finite-dimensional
state for the CV state. It exploits the fact \cite{caslav} that,
from the physical perspective, any two systems can be considered
as equivalent, if the probabilities for outcomes of all possible
future experiments performed on one and on the other are the same.
Mathematically, the requirement can be expressed as
\begin{equation}
\label{eq:equi}
\mbox{Tr}\left(\hat{A}(\vec{a})\hat{\rho}\right)=
\mbox{Tr}\left(\hat{\Omega}(\vec{a})\hat{\rho}_d\right)
\end{equation}
where $\hat{\rho}$ is the density matrix of any CV state while
$\hat{\rho}_d$ is that of a $d$-dimensional state.

Moreover, it is important to clarify that the mapping is a physically
possible quantum process. One can show that the observable
$\hat{A}(\vec{a})$ on a CV state results in a trace preserving, linear,
CP map $\varepsilon$ as
\begin{equation}
\hat{\rho}\rightarrow\hat{\rho}_d=\varepsilon(\hat{\rho})
\end{equation}
where $\hat{\rho}\in {\cal B}({\cal H})$ and $\hat{\rho}_d\in {\cal B}
({\cal K})$. We denote that ${\cal B}({\cal H})$ is the set of
operators defined in ${\cal H}$. Note also that ${\cal H}$ and ${\cal
  K}$ are the infinite and $d$-dimensional Hilbert spaces respectively.
In order to prove it, it is possible to make use of the correspondence
between the complete positive maps and positive-semidefinite operators
\cite{fiurasek}. The density matrix in ${\cal B}({\cal K})$ can be
expressed by the transformation $\varepsilon$ as follows
\begin{equation}
\label{eq:semi}
\hat{\rho}_d = \mbox{Tr}_{{\cal H}}\left(
\openone_{{\cal K}} \otimes
\hat{\rho}^{T}\hat{R}_{\varepsilon}\right)
\end{equation}
where $\hat{R}_{\varepsilon}$ is a positive-semidefinite operator
defined in ${\cal B}({\cal K}\otimes{\cal H})$. The
positive-semidefinite operator has the explicit form as
\begin{equation}
\hat{R}_{\varepsilon}=\sum_{n=0}^{\infty}\sum_{kl=0}^{d-1}|k\rangle\langle
l|\otimes|dn+k\rangle\langle dn+l|
\end{equation}
which satisfies the trace preserving properties of the CP map by
$\mbox{Tr}_{{\cal K}}(\hat{R}_{\varepsilon})=\openone_{{\cal H}}$.  The
correspondence between the CP map and the observable $\hat{A}(\vec{a})$
is confirmed from the dual map $\varepsilon^{\vee}$ of the map
$\varepsilon$ \cite{ariano} on the observable $\hat{\Omega}(\vec{a})$
and it is
\begin{eqnarray}
\varepsilon^{\vee}(\hat{\Omega}(\vec{a}))&\equiv&\mbox{Tr}_{{\cal
K}}\left(\hat{\Omega}(\vec{a})\otimes\openone_{{\cal H}}
\hat{R}_{\varepsilon}^{T_{{\cal H}}}\right)\nonumber\\
&=&\hat{A}(\vec{a})
\end{eqnarray}
where $T_{{\cal H}}$ denotes partial transposition on the Hilbert
space ${\cal H}$ only. We conclude that the measurement with the
observable $\hat{A}(\vec{a})$ on a CV state is equivalent to
consider the state as a $d$-dimensional state, which is mapped
from the CV state, with the measurement of
$\hat{\Omega}(\vec{a})$. It can also be said that the mapping is
a linear, trace preserving CP map from Eq.~(\ref{eq:semi}).

\subsection{Mapping of multi-mode state}
The mapping ${\cal B}({\cal H}_A\otimes{\cal H}_B)\rightarrow
{\cal B}({\cal K}_1\otimes{\cal K}_2)$ for a two-mode CV density matrix onto
a bipartite $d$-dimensional state is possible as
\begin{equation}
\label{eq:manymo} \hat{\rho}_{12}
=\mbox{Tr}_{AB}\left(\openone_{{\cal K}}\otimes\openone_{{\cal
K}}\otimes \hat{\rho}^{T}_{AB}\hat{R}_{A1}\hat{R}_{B2}\right).
\end{equation}
The mapping for an arbitrary number of modes can also be found as an
extension of Eq.~(\ref{eq:manymo}).

As an example, we consider a two-mode squeezed state $|\psi\rangle$
which can be generated by a non-degenerate optical parametric amplifier
\cite{loudon},
\begin{equation}
|\psi\rangle = \sum_{n=0}^{\infty} \frac{(\tanh r)^n}{\cosh
r}|n,n\rangle_{A,B}
\end{equation}
where $|n\rangle$ is a Fock state and $r$ is the squeezing
parameter. It is well-known that when squeezing parameter $r$
goes to infinity, the two-mode squeezed state approaches to the
EPR state \cite{EPR}. From Eq.~(\ref{eq:manymo}), one can map the
two-mode squeezed state onto the $d$-dimensional pure state:
\begin{equation}
|\psi_d\rangle = \frac{\mbox{sech} r }{\sqrt{1-\tanh^{2d}
r}}\sum_{n=0}^{d-1} (\tanh r)^n|n,n\rangle_{A,B}.
\end{equation}
The mapped state is a partially entangled pure state whose entanglement
is characterized by the squeezing parameter $r$. The state becomes
separable only when $r=0$ and it becomes maximally entangled for the
limit of $r\rightarrow \infty$.

\section{Numerical analysis based on SU($d$) group}

We investigate the optimal violation of the Bell's inequality
based on the Bell function $B_d$ in Eq.~(\ref{eq:bell}) for the
two-mode squeezed state. In order to search for optimization
values of the inequalities, we employ several numerical methods
such as steepest descent, conjugate gradient, and dynamic
relaxation. Each method has its own advantages and disadvantages
depending on the situations for optimization. The conjugate
gradient leads to rapid convergence for a nearly hyperbolic
function (where a bounded function looks like near its minimum).
The steepest descent method enables one to find persistently lower
values, even though it has disadvantages of slow convergence for
the nearly hyperbolic function that is squeezed in parameter
space. The dynamic relaxation method is in between the two
methods. We consider the dynamic relaxation method in detail as
the algorithms and implementations of the other methods can easily
be found in literatures \cite{Recipes}.

The dynamic relaxation method simulates a physical system under a
potential and a friction, which resembles the Car-Parrinelo method for
{\em ab initio} molecular dynamics \cite{Car85}.  Consider a bounded
function $B(\{p_i\})$ in terms of the parameter vector $p_i$. For an
optimization the method simulates a dynamic equation for a fictitious
classical particle, by regarding $p_i$ as its trajectory vector and
$B(\{p_i\})$ as a potential. The dynamic equation of motion can be
written as
\begin{eqnarray}
  \label{eq:eom}
  m \frac{d^2}{dt^2} p_i(t) = - \gamma \frac{d}{dt}p_i(t) - \frac{\partial}{\partial
  p_i} B(\{p_j\};t)
\end{eqnarray}
where $m$ is a mass of the fictitious particle and $\gamma$ is a
friction ratio. Note that the equation is a kind of the Langevin
equation. The particle will relax to the minimum of the potential.  The
solution to Eq.~(\ref{eq:eom}) approaches to the minimum of the function
$B(\{p_i\})$ exponentially due to the friction $\gamma$. A minimum is
claimed to be achieved when $ |\partial_{p_i} B(\{p_j\})| \times |p_i|
\leq 10^{-6}$. For the numerical implementation a Runge-Kutta method
is used to solve the dynamic equation with the following ranges of the
parameters: $m=0.1$, $\gamma \in (0.5, 1.5)$ and $\delta t \in (0.01,
0.1)$.  The maximum value of $B(\{p_i\})$ is obtained by replacing the
``potential'' $B(\{p_i\})$ with $-B(\{p_i\})$. For the optimizations of
Bell functions in Eq.~(\ref{eq:bell}), the parameter vector is given by
${\vec{a}} = (\vec{A}_1, \vec{A}_2, \vec{B}_1, \vec{B}_2)$ where
$\vec{A}_i$ and $\vec{B}_i$ are parameter vectors for unitary
transformations $\hat{U}(\vec{A}_i)$ and $\hat{U}(\vec{B}_i)$,
respectively, in the group SU($d$).
\vspace{0.5cm}
 \begin{figure}[htbp]
\begin{center}
   \leavevmode
   \epsfxsize=11 cm
   \epsfysize=9 cm
\epsfbox{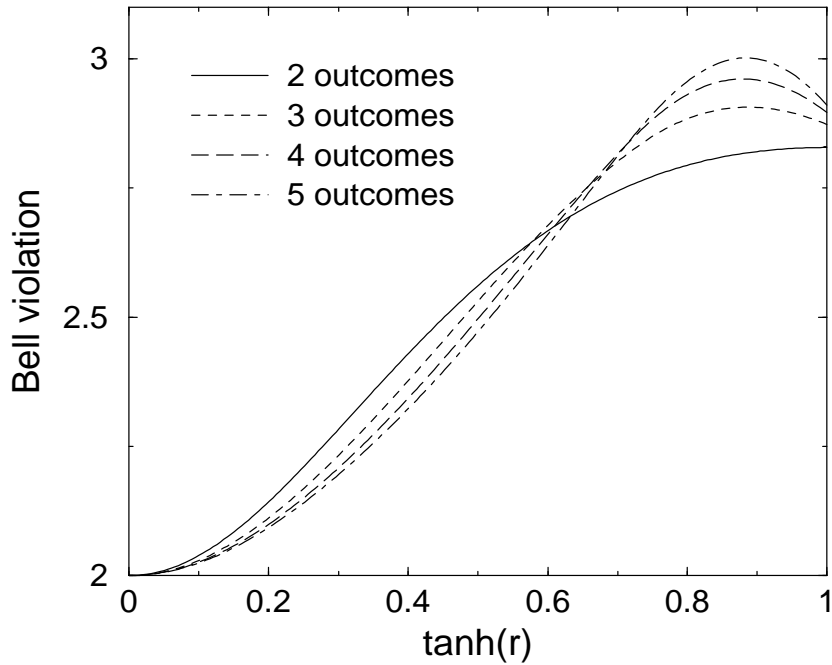}
\end{center}
\caption{\label{fig2} Violation of the Bell's inequality based on
the Bell function$B_d$ for two-mode squeezed state. Finite number
of measurement outcome are considered.}
\end{figure}

We optimize the value of the Bell function $B_d$ in
Eq.~(\ref{eq:bell}) with SU($d$) transformations for the two-mode
squeezed state. All the results are checked and they are
reproduced by conjugate gradient, steepest decent, and dynamic
relaxation methods. Fig.~\ref{fig2} presents the optimized value
of the Bell functions $B_d$ with respect to the strength of
squeezing $\tanh(r)$ for different numbers of measurement
outcome. The Bell function is upper bounded by 2, $B_d \leq 2$,
under the local realistic theory.

In Fig.~\ref{fig2}, we note that a two-mode squeezed state always
violates the inequality $B_d \leq 2$ for all $r>0$ regardless of
the number of measurement outcomes. Brukner {\em et al.}
calculated the values of the Bell function $B_d$ for $d=3$ based
on the QFT for a two-mode squeezed state. They do not always
achieve $B_d > 2$ for the squeezing parameter $r>0$ even though a
two-mode squeezed state is inseparable. Thus in order to properly
achieve the optimum value of the Bell function, we have to
consider all the possible transformations in SU($d$).
\begin{table}
\caption{\label{tab:1} The largest optimum values, $B_d(r_m)$,
  of the Bell function for two-mode
  squeezed state. $r_m$ is the squeezing parameter
  which maximized the value of
  $B_d$ for the given number, d, of measurement outcome.
  $B_d(\infty)$ is the value
  of $B_d$ for an infinite squeezing.}
\begin{indented}
\item[]
\begin{tabular}{@{\extracolsep{2.5cm}}cccc}
\br
$d$ & $r_m$ & $B_d(r_m)$ & $B_d(\infty)$ \\
\mr
2 & $\infty$ & 2.82843 & 2.82843 \\
3 & 1.407 & 2.90638& 2.87293 \\
4 & 1.373 & 2.96095 & 2.89624 \\
5 & 1.393 & 3.00187 & 2.91055\\
\br
\end{tabular}
\end{indented}
\end{table}

On the other hand, in the limit of $r \rightarrow \infty$, a two-mode
squeezed state becomes a regularized EPR state which is mapped onto a
maximally entangled state in finite dimensional Hilbert space. However,
for the maximally entangled state, the QFT suffices to obtain the
optimum value of $B_d$ as we have already discussed.

For a given number of measurement outcomes, the amount of violation
increases and decreases with respect to the squeezing parameter $r$. Let
$r_d$ denote the value of squeezing parameter that gives the largest
violation for a given measurement with $d$ number of outcomes. As shown
in Table~\ref{tab:1}, except for the dichotomic measurement, the
infinitely squeezed state violates the inequality less than some
partially entangled states. As increasing the number of outcomes, the
largest optimum values of $B_d$ monotonically increase.  It is also
found in Fig.~\ref{fig2} that for the high squeezing regime the higher
number of outcomes gives stronger violation while the result is reversed
for the small squeezing regime.

\section{Final remarks}

We studied the most general $d$-outcome measurement for the Bell's
inequalities of a bipartite system. In order to construct the
inequalities, we introduced a classically correlated observable
which is constructed in terms of local measurements and classical
communications. For the configuration of the local measurements,
we considered general transformations in SU($d$). It was found
that the number of parameters for the nontrivial operation is
reduced to $(d^2-d)$.  After inspection of symmetries, we derived
the Bell function that is composed of the correlation functions.
This Bell function is equivalent to that found by
Collins-Gisin-Linden-Massar-Popescu \cite{col}. The present
numerical analysis shows that, when the system is in a maximally
entangled state, the QFT is an optimal transformation for each
local measurement. However, we show that this does not hold when
the system is in a partially entangled state.

In order to utilize the CV state for the nonlocality test, we
investigated the mapping between a CV state and arbitrary
dimensional system which was devised by Brukner \textit{et
al.}\cite{caslav}. We found the mapping is linear, trace
preserving and complete positive map and it corresponds to a
highly degenerate $d$-outcome measurement on a CV state. By
applying the highly degenerate measurements, we investigated the
optimal violation of the Bell's inequality for the two-mode
squeezed state.  Regardless of the degree of squeezing and the
number of outcomes, the two mode squeezed state always violates
the Bell's inequalities. This opens a possibility to extend
Gisin's theorem \cite{gisin91} states that a pure entangled
bipartite system always shows nonlocality not only for the case of
dichotomic measurement but also for the case of a measurement with
an arbitrary number of measurement outcomes.

\ack This work was partially supported by the UK Engineering and
Physical Science Research Council and by the Korea Research
Foundation Grant (KRF-2003-070-C00024). W. Son
 acknowledges the
ORS award scheme for the financial support. JL was supported by
the KOSEF through the Quantum Photonic Science Research Center and
the Post-doctoral Fellowship Program of the KOSEF.

\end{document}